\documentclass[aps,prl,showpacs,twocolumn,letterpaper]{revtex4}
\usepackage{amsmath,graphicx,amssymb}

\newcommand{\ha}{\hat{a}}
\newcommand{\hx}{\hat{x}}
\newcommand{\hy}{\hat{y}}

\begin{document}
\title{Fractional quantum Hall states of atoms in optical Lattices}

\author{Anders S. S\o rensen$^{1,2}$, Eugene Demler$^2$, and Mikhail
D. Lukin$^{1,2}$}
\affiliation{$^1$ITAMP, Harvard-Smithsonian Center for Astrophysics and \\
$^2$Physics department,  Harvard university,
Cambridge Massachusetts 02138
}  

\begin{abstract}
We describe a method to create fractional quantum Hall states of
atoms confined   
in optical lattices. We show that the dynamics of the atoms in the
lattice is analogous to  the motion of a charged 
particle in a magnetic field if  an oscillating quadrupole
potential is applied together with  a periodic modulation of the
tunneling between lattice sites. We demonstrate that in a suitable
parameter regime the ground state 
in the lattice is of the fractional quantum Hall type and we show 
how  these states can be reached by melting a Mott insulator state 
in a super lattice potential. Finally we discuss 
techniques to observe these strongly
correlated states.

  \end{abstract}


\maketitle

Ultra-cold atomic gasses \cite{BEC-general}
 provide a unique
access to  quantum  many body systems with well understood and controllable
interactions. Whereas most of the experiments in this field
have been carried  out in the regime of 
weak interactions,  the recent achievements involving Feshbach resonances
\cite{feshbach} and the realization of a Mott-insulator
state of   
atoms in optical lattices \cite{bloch,esslinger}  enters into the
regime of strong    
interaction with a richer and more complex many body dynamics. 
At the same time a realization of strongly correlated states of 
fractional quantum Hall type
\cite{quantumhall}  has recently been 
suggested in cold atomic gases \cite{rotating}. These
proposals involve atoms in rotating harmonic traps, which  
mimic the effective magnetic field. However, weak interaction between
the particles (and correspondingly small gap in the excitation
spectrum), required precision on trap rotation and finite temperature
effects make 
these proposals difficult to realize experimentally.  In this
Letter we present a novel method that uses atoms in optical lattices
to create states of the fractional  quantum Hall type.
  Since the interactions of atoms localized in the
lattices  are strongly enhanced compared to the interaction of atoms
in free space, these states are characterized by large energy gaps.


The fractional quantum Hall effect occurs for electrons confined to a
two dimensional plane (the $xy$ plane) in a strong magnetic field. In
the simplest form the effect occurs if
the number of magnetic fluxes $N_\phi$ (measured in units of the
fundamental flux quanta $\Phi_0=2\pi\hbar/e$) is an integer $m$ times the
number of particles $N_\phi=m\cdot N$. At this value of the magnetic
field the ground state of the system is an incompressible quantum
liquid which is separated from all other states by an energy gap and 
is well described by the Laughlin wavefunction \cite{laughlin}
\begin{equation}
  \label{eq:laughlin}
  \Psi(z_1,z_2,....z_N)={\rm e}^{{\left(-\sum_j |z_j|^2/4\right)}}
  \prod_{j<k}(z_j-z_k)^m, 
\end{equation}
where $z=x+i y$, and where we have assumed the symmetric gauge and
suitable magnetic units. Due to the Pauli principle
only the states with odd (even) $m$
is applicable to fermions (bosons). In this Letter we for simplicity only 
consider bosons and $m=2$.
 
Below we describe  a method to produce the states 
(\ref{eq:laughlin}) for a system of ultra-cold atoms in an optical
lattice. 
In such a system these states are protected by an excitation gap
which is
controlled by  the tunneling energy. 
For typical experimental parameters  this is much larger than the
energy scale in the macroscopic magnetic traps considered previously
\cite{rotating}.  
 The larger energy gap is a
clear advantage from an experimental point of view because the state
is more robust to external perturbations, and the present approach
could therefore enable the realization  of the Laughlin state with a
larger number of particles.  We note further that experiments with
periodically modulated quantum Hall probes \cite{geisler}  and tunnel
coupled superconducting islands \cite{josephson} are trying to reach
the regime studied here, and the extension  of the quantum Hall physics to
a lattice system is therefore an interesting
non-trivial problem in its own right.
For non-interacting particles 
the energy spectrum on the lattice (the so-called Hofstadter
butterfly \cite{hofstadter})
is very
different from that of Landau levels. We nevertheless show that  
at least in a certain parameter regime, the Laughlin state
provides a reasonable description for the resulting many-body physics 
of the strongly interacting system. 

We consider atoms trapped in a square two-dimensional optical
lattice.
 If the atoms in the lattice
are restricted to the lowest Bloch-band the system
can be described by a Bose-Hubbard Hamiltonian \cite{jaksch98}
\begin{equation}
  \label{eq:bosehubbard}
  H=-J \sum_{\{ j,k\} } (\ha^\dagger_j\ha_k+\ha^\dagger_k\ha_j )
   +U \sum_j  n_j(n_j-1),  
\end{equation}
where the first sum is over neighboring sites $j$ and $k$, $\ha_j$ and
 $n_j=\ha^\dagger_j\ha_j$ are
the boson annihilation and number operators  on site $j$, $J$ is the tunneling
amplitude, and $U$ is the onsite interaction energy.
The individual lattice sites will below
be specified by
a pair of integers $x$ and $y$. 

\begin{figure}
\includegraphics[width=8cm]{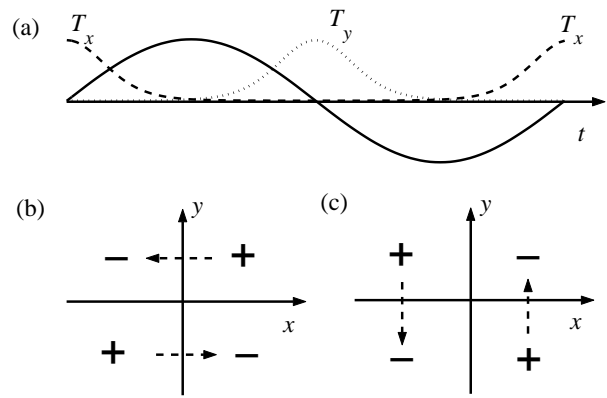}
\caption[]{(a) One period of the sequence used to create an effective magnetic
field. The time evolution of the quadrupole potential is shown by the
full curve and the dashed and dotted lines indicate the tunneling in
the $x$ and $y$ direction. For illustration the shown 
evolution of the
tunneling is obtained from a sinusoidal variation of the lattice
potential between 5 and 40 recoil energies \cite{jaksch98}. (b) and
(c) Physical explanation of the procedure. (b) Tunneling in the
$x$-direction is followed by a positive potential in the first and
third quadrant (signs in the figure), and hence atoms will experience
a lower potential by moving in the
direction of the dashed arrows.
 (c) Same as (b) but with  tunneling in
the $y$-direction and opposite sign
of the potential. When combined the
dashed lines in (b) and (c) makes a circular cyclotron motion. }   
\label{fig:mag}
\end{figure}

An essential ingredient in our proposal is  an
effective magnetic field for  neutral atoms in optical lattices. 
Different approaches which attain this goal has already been proposed
\cite{magnetic}, but here we present an  
alternative  procedure that may simplify the experimental realization.
Our procedure involves a combination of a time-varying quadrupolar
potential $V(t)=V_{\rm qp}
\sin(\omega t) \cdot \hx\cdot \hy$, and a modulation of the
tunneling in time. The tunneling between neighboring sites decreases
exponentially with the intensity of the lasers creating the
lattice, whereas the shape of the wavepacket (the Wannier functions)
has a much weaker dependence \cite{jaksch98}.
By varying the
laser intensity the tunneling can therefore be varied rapidly in
time. Assume that the tunneling in the $x$-direction is
turned on for a short period around $\omega t=\pi\cdot 2
n$, ($n=0,1,2...$) and that the tunneling in the $y$-direction is
turned on for a short period around  $\omega t=\pi\cdot(2
n+1)$ (see Fig.\ \ref{fig:mag}
(a)). As illustrated by the simplified picture  in Fig.\
\ref{fig:mag} (b) and 
(c) such a time sequence creates an effective Lorentz force (magnetic
field) in the lattice. This can be shown mathematically  by
assuming that the tunneling is only present in a very short time
interval $\tau$. 
The total evolution after $m$ periods is then given by
(neglecting for now the interaction between the particles)
\begin{equation}
  \label{eq:U}
\begin{split}
  U{\left(t=\frac{m 2\pi}{\omega}\right)}=\Big(&  {\rm e}^{-i\tau
  T_x/2\hbar} {\rm e}^{
  2iV_{\rm qp}\hx\hy/\omega\hbar}  {\rm e}^{-i\tau
  T_y/\hbar}\times \\
&  {\rm e}^{-2iV_{\rm qp}\hx\hy/\omega\hbar}  {\rm e}^{-i\tau
  T_x/2\hbar}  \Big) ^m,
\end{split}
\end{equation}
where $T_x$ and $T_y$ are the kinetic energy operators describing
tunneling in the $x$ and $y$ direction respectively.
By using $\exp(i2\pi \alpha \hx\hy)\ha_{x,y+1}^\dagger \ha_{x,y}
\exp(-i2\pi\alpha \hx\hy) = \ha_{x,y+1}^\dagger \ha_{x,y}  \exp(i2\pi\alpha
x)$ with $\alpha=V_{\rm qp}/\pi\hbar \omega$ 
 this expression can be reduced  
to the time evolution $U=\exp(-iH_{{\rm eff}}t/\hbar)$ from an effective
Hamiltonian 
\begin{equation}
  \label{eq:ham}
  H_{{\rm eff}}\approx -J 
  \sum_{x,y} \ha_{x+1,y}^\dagger
 \ha_{x,y}+
 \ha_{x,y+1}^\dagger \ha_{x,y} {\rm e}^{i2\pi\alpha
  x}+{\rm H.C.}
 ,
\end{equation}
where the tunneling strength $J$ is the average tunneling per period,
and where we have omitted
terms of order 
$J (J/\omega)^2$. Eq.\ (\ref{eq:ham}) 
describes the behavior
of a charged particle on a lattice  with a magnetic
flux $\alpha\Phi_0$ going through each unit cell, and
hence the procedure introduces an 
 effective magnetic field in the lattice. The gauge in  Eq.\ 
(\ref{eq:ham}) (Landau gauge) is determined by the time we terminate
the sequence in Fig. \ref{fig:mag}
(a) and a different gauge would appear if we terminated at a
different time. 


 We now turn to the fractional quantum
Hall effect for strongly interacting atoms in the presence of an
effective magnetic
field. In the limit of small $\alpha$ and a small number of 
atoms per lattice site the dynamics of the system reduces to the
continuum limit of particles in a magnetic field with infinitely
short range interactions. If the interaction is repulsive,  the Laughlin
wavefunction (\ref{eq:laughlin}) is known to be  the absolute
ground state of the system when $N_\phi=m\cdot N$
\cite{haldane-ground}. This limit corresponds to a very low density
gas and  therefore the energy gap between the ground and excited
states is vanishingly small (see below). To extend this analysis
to the situation  when $\alpha$ is no longer vanishingly small we have 
performed a direct numerical diagonalization of the Hamiltonian in
Eq.\ (\ref{eq:ham}) for a small number of hard-core bosons
(corresponding to $J\ll U$). (In 
different contexts, similar problems were considered in
Ref. \cite{canright}).

\begin{figure}[bt]
  \centering
  \includegraphics[width=8cm]{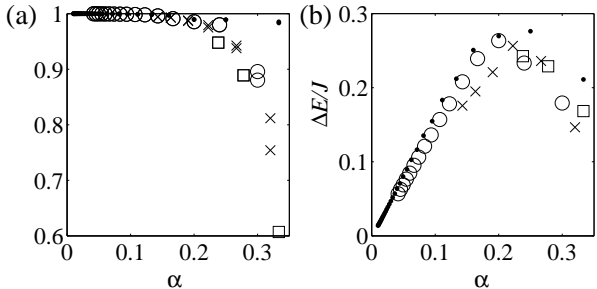}
  \caption{(a) Overlap of the ground state wavefunction with the Laughlin
  wavefunction (\ref{eq:laughlin}). (b) Energy gap $\Delta E$ to
  the lowest
  excited state. In the figure we have fixed the number of
  particles and fluxes ($N_\phi=2N$) and vary the flux per
  unit cell $\alpha$ by varying the size of the lattice. The shown results
  are for $N=2$  ($\bullet$), $N=3$ ($\circ$), 
  $N=4$ ($\times$), and $N=5$  ($\Box$).}  
  \label{fig:overlap}
\end{figure}

\begin{figure}[t]
  \centering
  \includegraphics[width=8cm]{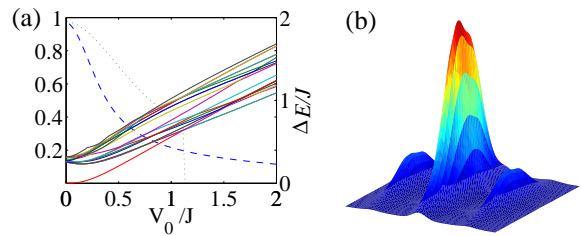}
  \caption{(a) Creation of the Laughlin state by melting a Mott
  insulator. Full lines (right axis) energy relative to the
  ground state of the 19 lowest excited states 
  for different amplitudes ($V_0$) of the super 
  lattice potential.  
  Dashed (dotted) line overlap of the ground state
  (first excited state) with the Laughlin wavefunction
  (\ref{eq:laughlin}) (left axis). Note that at $V_0=0$ there are two
  nearly degenerate states because of the  
  combination of magnetic field and periodic boundary
  conditions. These two states have a 98\% overlap with the two
  possible Laughlin states.
  Results are for a $6\times 6$
  lattice with 4 atoms and 8 fluxes, $\alpha\approx 0.22$. The periods
  of the super lattice are $p_x=p_y=3$. (b) Density distribution after
  expansion from the lattice (arbitrary
  units). For illustration we have assumed a lattice depth of  10
  recoil energies \cite{jaksch98} and $\alpha\approx0.22$ as in (a).}
  \label{fig:energy}
\end{figure}

To investigate the effect of finite $\alpha$ we fix the number of
fluxes and particles so that $N_\phi=2N$ and vary $\alpha$ by changing
the size of the 
lattice. 
(We only consider lattices where the size in the $x$ and $y$ 
direction are the same or differ by unity).
In Fig.\ \ref{fig:overlap} (a) we show the overlap of
the ground state wavefunction from the diagonalization of the
Hamiltonian  with the Laughlin
wavefunction (\ref{eq:laughlin}). 
For $\alpha\lesssim 0.3$ the ground state has a
very good overlap with the Laughlin wavefunction. We have assumed periodic
boundary conditions in order to represent the bulk properties of 
a large optical lattice. For the situation considered here ($m=2$) 
 the combination of
periodic boundary conditions and a magnetic field gives rise to a two
fold degeneracy of the ground state in the continuum limit
($\alpha \ll 1$), where the two ground states only differ by
their center of mass wavefunctions \cite{haldane-sym}.
 The symmetry analysis leading
to this degeneracy does not apply in the presence of the lattice (for
a discussion of the symmetries in a periodic potential see Ref.\
\cite{lattice-sym}).  For all points in Fig. \ref{fig:overlap} (except
the point
$N=5$, $\alpha=1/3$), however, the diagonalization gives two 
almost degenerate ground states 
which are separated from the 
excited states (see Fig.\ \ref{fig:energy} (a) at $V_0=0$).
 The periodic generalization of the Laughlin
wavefunction \cite{haldane-periodic} also have a two-fold center of
mass degeneracy and in Fig.\ \ref{fig:overlap} (a) we show the overlap of
the two lowest states from the diagonalization with the subspace
spanned by the 
Laughlin wavefunctions. 

Our simulations show  a very good overlap with the
Laughlin wavefunction for $\alpha\lesssim 0.3$, but the overlap start
to fall off for $\alpha\gtrsim 0.3$. We emphasize that the excellent
overlap with the Laughlin wavefunction is not accidental, e.g., for
$N=5$ and $\alpha \approx 0.24$ the size of the Hilbert space is $8.5\cdot
10^5$ and the overlap is 95\%. These numerical calculation 
therefore provide strong 
evidence that the Laughlin wavefunction captures the essential
properties of the many-particle system.  Experimentally it is
desirable to have as large 
an excitation gap as possible. From the results in Fig.\
\ref{fig:overlap} (b) we see that the optimal regime is $\alpha\sim
0.2$ (although the data still have finite size effects).
In this region the Laughlin
wavefunction (\ref{eq:laughlin}) is a very good description of the state.  

In order to reach the Laughlin state experimentally it is  necessary
to adiabatically 
load a cold Bose-Einstein condensate (BEC) into the lattice. 
We expect the transition between the superfluid BEC and fractional
quantum Hall states to be a direct first order phase transition or
proceed via several intermediate phases \cite{hierarchy}. Thus, 
a direct transition between the BEC and the fractional quantum Hall state
is likely to create many excitations. 
To avoid this problem we suggest to accomplish the Laughlin state
preparation through a Mott-insulator state 
\cite{jaksch98,bloch,esslinger}. Such a state  with an integer number of atoms
per site can be reached from a BEC by adiabatically  raising the
lattice potential so that $U$ in Eq.\ (\ref{eq:bosehubbard}) is much
larger than $J$. We consider a  situation where an 
additional (weak) super lattice $V=V_0(\sin^2(\pi x/p_x)+\sin^2(\pi
y/p_y))$ is present, as realized experimentally in Ref.\
\cite{superlattice}. By loading a BEC into the combined 
potential it is then possible to reach a Mott insulator state with a
single atom at each of the potential minima of the
super lattice. In this state the atoms cannot move and are
hence unaffected by the turn on of the effective magnetic field. By
reducing the super lattice potential it is then possible to
adiabatically reach the Laughlin state, as demonstrated in Fig.\
\ref{fig:energy} (a). In the figure we show the evolution of the lowest
energy levels when we reduce the strength of the super lattice
potential $V_0$. For large values of $V_0$ the ground state is the
Mott-insulator state which is well separated from all excited
states. When $V_0=0$ there are two nearly degenerate ground states
because of the center of mass degeneracy mentioned above. Apart from
this mathematical artifact of the periodic boundary conditions,
the ground state
is always well separated from the excited states by an energy
gap, and the ground state of the system smoothly changes into the Laughlin
state, so that it is possible to reach the Laughlin state by
adiabatically reducing the super lattice potential. 

We next consider experimental issues involving the realization and
detection of quantum Hall states in the lattice. 
Under realistic conditions the required "hard-core" limit can be
reached, e.g., in atomic Rb with a
tunneling rate $J/2\pi\hbar$ of the order of hundreds of Hz
\cite{bloch,jaksch98}, which 
indicates that energy gaps in the range 10-100 Hz can be
obtained. To observe the quantum Hall states the temperature of the atoms
need to below this excitation gap which requires realistic 
temperatures of a few
nK.  A limitation
for the practical realization of the present proposal is that the
oscillating potentials produce strong phase shifts on the atoms. If there
is a total of $N_\phi$ fluxes in the lattices the phase shift on atoms
in the outermost regions is on the order of $N_\phi$ per
half-cycle. Hence a practical implementation requires strong
gradients and it is necessary to ensure that the phase shift in one
half of the pulse exactly balances the phase shift in the other.
Another limitation is that the oscillating quadrupole potential can
excite  higher Bloch bands. If we approximate the wells by a harmonic 
potential the weight on the excited state in the outermost regions of
the lattice is $w_{{\rm ex}}\sim \pi
N_\phi\alpha (a_0/\lambda)^2 \omega^2/\nu_b^2$, where $\nu_b$ is the
Bloch band separation, and where the
ground state width $a_0$ is of order $\lambda/10$ for typical lattice
parameters \cite{jaksch98}. With $\omega\sim \nu_b/10$ 
this is not a major concern for $N_\phi \lesssim 10^2 - 10^3$. 

To demonstrate experimentally that one has reached the Laughlin
states one would ideally probe some of the unique features of the state,
such as incompressibility, the fractional charge of the excitations or
their anyonic character. Such probes will most likely be very
challenging to implement and we shall now discuss a simpler
experimental indication. In most experiments with cold trapped atoms
the state is probed by releasing the atoms from the trap and imaging
the momentum distribution. 
To find the results of
such an expansion we use the continuum wavefunction
(\ref{eq:laughlin}) which 
is a good 
description in the regime we are interested in.   In the
lowest Landau level the single particle density
matrix for any state with constant density was found in
Ref. \cite{densitymatrix}.
From this density matrix we find
the asymmetric expansion shown in Fig. \ref{fig:energy} (b).
This density distribution is clearly
distinct from, e.g., a superfluid state which will have Bragg
peaks, and a Mott-insulator, which gives a symmetric
distribution \cite{bloch}.
This method, however, does not reveal detailed information
about the state except that it is in the lowest Landau level. 
Further insight 
can be obtained by
measuring higher order correlation 
functions  \cite{read2003,ehud-expansion} or by measuring the
excitation spectrum through stimulated Bragg scattering \cite{bragg}.

To summarize, we have presented a feasible method to construct fractional
quantum Hall states in an optical lattice. Compared to
previous proposals  with cold atoms
\cite{rotating}  the
optical lattice approach results in a more robust quantum Hall 
state, since it is protected by a larger gap. The present approach  
therefore reduces the experimental requirement and could facilitate the
observation of such states with a larger number of particles. 

Several interesting new avenues are opened by this work.
  First of all,  it would be interesting to
understand the exact nature of the ground state  when 
there is a large flux fraction per unit
cell $\alpha\gtrsim 0.3$. 
We have not made any modification to the Laughlin
wavefunction to take the lattice into account, and one would expect
this to reduce the overlap in Fig.\ \ref{fig:overlap} (a) when
$\alpha$ is not vanishingly 
small. The decrease in the overlap is, however, much less abrupt for the
single particle wavefunctions. We therefore do not expect that this
can explain the observed results. Alternatively the decrease
could be caused by the system entering a different phase 
for $\alpha\gtrsim 0.3$, e.g., 
a superfluid state with a vortex lattice. This possibility is 
supported by the 
decrease in the excitation gap above  $\alpha\gtrsim 0.25$, see Fig.\
\ref{fig:overlap} (b). We further observe an increase in the largest
eigenvalue of the one-particle density matrix above $\alpha\gtrsim0.3$
(especially for $\alpha=1/3$ and $\alpha=1/2$ that correspond to
a commensurate density of vortices) which is consistent with
a superfluid state. We cannot, however, draw any definite conclusion
from our present numerical results.
In addition the present method for creating the quantum Hall state in
the lattice can be easily extended to yield different magnetic field for
different internal 
atomic states of multi-component bosons. Using this approach
effective non-abelian gauge fields can be created. Therefore the
present method may allow one  to explore experimentally the novel properties
of a many particle 
system in the presence of such a field.
 
We are grateful for useful discussion with E.\ Altman, M.\ Greiter,
B.I.\ Halperin, P.\ Zoller,  
and the Quantum optics group at
ETH, Z\"urich.
This work was supported 
by NSF  through grants PHY-0134776 and
DMR-0132874 and through the grant to ITAMP,  by the Danish Natural
Science Research Council, and by the 
Sloan and Packard foundations.



\begin{thebibliography}{99}
\bibitem{BEC-general} J.R.\ Anglin and W.\ Ketterle, Nature {\bf 416},
  211 (2002).

\bibitem{feshbach} C.A.\ Regal, M.\ Greiner, and D.S.\ Jin,
  Phys.\ Rev.\ Lett.\ {\bf 92}, 040403 (2004); M.W.\ Zwierlein {\it et
  al.},
{\it ibid.} {\bf 92}, 120403 (2004); M.\ Bartenstein {\it et
 al.}, {\it ibid.} {\bf 92}, 120401 (2004);  J.\ Kinast {\it et al.},
 {\it ibid.} {\bf 92}, 150402 (2004); T.\ Bourdel {\it et al.},
cond-mat/0403091. 

\bibitem{bloch} M.\ Greiner, O.\ Mandel, T.\ Esslinger, T.W.\ H\"ansch and
    I.\ Bloch,  Nature {\bf 415}, 39 (2002).

\bibitem{esslinger} 
T.\ St{\"o}ferle {\it et al.}, 
 Phys.\ Rev.\ Lett.\ {\bf 92}, 130403 (2004).


\bibitem{quantumhall} {\it The quantum Hall effect}, edited by
   R.E.\ Prange and S.M.\ Girvin (Springer-Verlag, New York, 1990).

\bibitem{rotating} N.K.\ Wilkin and J.M.F.\ Gunn, Phys.\ Rev.\ Lett.\
  {\bf 84}, 6 (2000); T.-L.\ Ho,  {\it ibid.} {\bf 87} 060403 (2001);
  B.\ Paredes, P.\ Fedichev, J.I.\ Cirac, and
  P. Zoller,  {\it ibid.} {\bf 87}, 010402 (2001). 

\bibitem{laughlin} R.B.\ Laughlin, Phys.\ Rev.\ Lett.\ {\bf  50}, 1395 (1983).

\bibitem{geisler}  M.C.\ Geisler {\it et al.}, Phys.\ Rev.\ Lett.\
{\bf 92}, 256801 (2004).  

\bibitem{josephson} H.S.J. van der Zant {\it et al.}, Phys.\ Rev.\
Lett.\ {\bf 69},  2971 (1992); B.\ Pannetier {\it et al.}, {\it ibid.}
{\bf 53}, 1845 (1984).

\bibitem{hofstadter} D.R.\ Hofstadter, Phys.\ Rev.\ B {\bf 14}, 2239 (1976).
\bibitem{jaksch98} D.\ Jaksch, C.\ Bruder, J.I.\ Cirac, C.W.\
Gardiner, and P.\ Zoller, Phys.\ Rev.\ Lett.\ {\bf 81}, 3108 (1998).

\bibitem{magnetic} D.\ Jaksch and P.\ Zoller, New J.\ Phys.\ {\bf 5},
56 (2003); E.J. Mueller, cond-mat/0404306.   

\bibitem{haldane-ground} F.D.M.\ Haldane, Phys.\ Rev.\ Lett.\ {\bf
51}, 605 (1983). 

\bibitem{canright}
G.S.\ Canright, S.M.\ Girvin, and A. Brass, Phys. Rev. Lett {\bf 63},
2291 (1989); M.\ Niemeyer, J.K.\ Freericks, and H.\ Monien,
Phys.\ Rev.\ B {\bf 60}, 2357 (1999). 

\bibitem{haldane-sym} F.D.M.\ Haldane, Phys.\ Rev.\ Lett.\ {\bf 55},
  2095 (1985).

\bibitem{lattice-sym} A.\ Kol and N.\ Read, Phys.\ Rev.\ B {\bf 48},
  8890 (1993). 

\bibitem{haldane-periodic} F.D.M.\ Haldane and  E.H.\ Rezayi,
 Phys. Rev. B {\bf 31}, 2529 (1985). 

\bibitem{hierarchy}
N. Read, Phys.\ Rev.\ Lett.\ {\bf 65}, 1502 (1990);
S. Kivelson, D.H. Lee, S.C. Zhang, Phys.\ Rev.\ B {\bf 46}, 2223 (1992);
X.G. Wen, Advances in Physics {\bf 44}, 405 (1995).


\bibitem{superlattice} S.\ Peil {\it et al.},
 Phys.\ Rev.\ A {\bf 67}, 051603 (2003). 



\bibitem{densitymatrix} A.H\ MacDonald and S.M.\ Girvin, Phys.\ Rev.\
  B {\bf 38} 6295 (1988). 





\bibitem{read2003}
N.\ Read and N.R.\ Cooper, Phys. Rev. A {\bf 68}, 35601 (2003).

\bibitem{ehud-expansion} E.\ Altman, E.\ Demler, M.D.\
   Lukin, cond-mat/0306226.

\bibitem{bragg} J.\ Stenger {\it et al.}, Phys.\ Rev.\ Lett.\ {\bf82},
4569 (1999).
\end{thebibliography}
\end{document}